\documentstyle[12pt]{article}
\begin{document}
\renewcommand{\thefootnote}{\fnsymbol{footnote}}
\begin{center}
\begin{LARGE}
The Nolen-Schiffer anomaly and \\
isospin symmetry breaking in nuclear matter
\end{LARGE}
\end{center}
\vspace{1cm}
\begin{center}
\begin{large}
K.~Saito\footnote{ksaito@nucl.phys.tohoku.ac.jp} \\
Physics Division, Tohoku College of Pharmacy \\ Sendai 981, Japan \\
and  \\
A.~W.~Thomas\footnote{athomas@physics.adelaide.edu.au} \\
Department of Physics and Mathematical Physics \\
University of Adelaide, South Australia, 5005, Australia
\end{large}
\end{center}
\vspace{1cm}
\begin{abstract}
The quark-meson coupling model which we have developed previously
is extended to incorporate the $\delta$ meson. It is then used to
study the
Nolen-Schiffer anomaly and isospin symmetry breaking in nuclear
matter.  We find that, in combination with the {\it u-d\/} mass
difference,
the difference between quark scalar densities
in protons and neutrons generates an effective neutron-proton mass
difference of the right magnitude. Finally we find that
isospin
symmetry breaking in the quark condensates
can be directly related to the $\delta$ meson field.
\end{abstract}
%
%
\newpage
The Nolen-Schiffer (NS) anomaly~\cite{ok,ns} (sometimes called the
Okamoto-Nolen-Schiffer anomaly) is a long-standing problem in nuclear
physics.  The anomaly is the discrepancy between experiment and
theory
for the binding energy differences of mirror nuclei.
It is conveniently expressed in terms of the quantity $\Delta E_c$:
\begin{equation}
\Delta E_c = M_{Z_{>}} - M_{Z_{<}} + \delta_{np} + E_c, \label{eq:def}
\end{equation}
where $M_{Z_{>}(Z_{<})}$ is the atomic mass of the larger (smaller)
charge
nucleus, $\delta_{np} = 0.782$ MeV is the neutron-proton atomic
mass difference and $E_c$ is the Coulomb correction associated with
the additional proton in $Z_>$.  Conventional nuclear contributions
to the
anomaly are thought to be at the few per
cent level and cannot explain the experimental findings~[1-4].
The effects of charge symmetry breaking in
the nuclear force~\cite{ok,hs}, especially $\rho$-$\omega$ mixing,
seem to reduce the discrepancy~[4-6].
However, recent investigations of
the off-shell variation of the $\rho$-$\omega$ mixing amplitude
have put this explanation into
question~\cite{ght,hat}.  In an alternate approach, using the
Nambu--Jona-Lasinio (NJL) model, Henley and Krein~\cite{hk} have
indicated that the anomaly might be related to the partial
restoration of chiral symmetry in nuclear matter. More recent
theoretical investigations have involved QCD
sum-rules~\cite{hb}.
\par
In this letter we shall extend the quark-meson coupling (QMC)
model~\cite{sath} to incorporate an isovector, scalar meson, the
$\delta$.
We then use it to investigate
the NS anomaly and isospin symmetry breaking in matter.  In this
model, which is a natural development of the earlier work of
Guichon~\cite{guichon} and Yazaki et al.~\cite{yazaki},
nuclear matter consists of non-overlapping nucleon
bags bound by the self-consistent exchange of $\sigma$, $\omega$,
$\rho$ and $\delta$ mesons in the mean-field approximation (MFA).
It provides an excellent description of the properties of both
nuclear
matter and nuclear structure
functions~\cite{sath,saito}.  Furthermore, the relationship between
the QMC model and QHD~\cite{serot} has been clarified~\cite{sath}.
\par
Let the mean-field values for the $\sigma$, $\omega$ (the time
component), $\rho$ (the time component in the third direction of
isospin) and $\delta$ (in the third direction of isospin) fields,
in uniformly distributed nuclear matter with $N \ne Z$, be
$\bar{\sigma}$, $\bar{\omega}$, $\bar{b}$ and $\bar{\delta}$,
respectively.  The
nucleon is described by the static spherical MIT bag in
which quarks interact (self-consistently)
with those mean fields.  The Dirac equation for
a quark field, $\psi$, in a bag is then given by
\begin{equation}
[i\gamma\cdot\partial - (m_i - V_{\sigma} - \frac{1}{2}
\tau_zV_{\delta}) - \gamma^0(V_{\omega} + \frac{1}{2}\tau_zV_{\rho})]
\psi_{i/j} = 0, \label{eq:dirac}
\end{equation}
where $V_{\sigma}=g_{\sigma}^{q}\bar{\sigma}$, $V_{\omega}=
g_{\omega}^{q}\bar{\omega}$, $V_{\rho}=g_{\rho}^q\bar{b}$ and
$V_{\delta}=g_{\delta}^q\bar{\delta}$ with the quark-meson coupling
constants, $g_{\sigma}^q$, $g_{\omega}^q$, $g_{\rho}^q$ and
$g_{\delta}^q$.  The subscripts, {\it i\/} and {\it j\/}, denote
the {\it i}-th quark in the proton or neutron ({\it j\/}={\it p\/}
or {\it n\/}).  Here we deal with {\it u\/} and {\it d\/} quarks
({\it i\/}={\it u\/} or {\it d\/}) only.  The bare quark mass is
denoted by $m_i$ and
$\tau_z$ is the third Pauli matrix.  The normalized,
ground state
for a quark in the nucleon is then given by
\begin{equation}
\psi_{i/j}(\vec{r},t) = {\cal N}_{i/j} \exp[-i\epsilon_{i/j} t/R_j]
{j_{0}(x_ir/R_j) \choose
i\beta_{i/j} {\vec{\sigma}}\cdot\hat{r}j_{1}(x_ir/R_j)}
{\frac{\chi_i}{\sqrt{4\pi}}}, \label{eq:psiq}
\end{equation}
where
\begin{equation}
\epsilon_{i/j} = \Omega_{i/j} + R_j(V_{\omega} \pm \frac{1}{2}
V_{\rho}),
\mbox{ for } {u \choose d} \mbox{ quark} \label{eq:epq}
\end{equation}
\begin{equation}
{\cal N}_{i/j}^{-2} = 2R_j^3j^2_0(x_i)[\Omega_{i/j}(\Omega_{i/j} - 1)
+
R_jm_i^{\star}/2]/x_i^2, \label{eq:norm}
\end{equation}
\begin{equation}
\beta_{i/j} = \sqrt{(\Omega_{i/j} - R_jm_i^{\star})/(\Omega_{i/j} +
R_jm_i^{\star})}, \label{eq:betq}
\end{equation}
with $\Omega_{i/j} = \sqrt{x_i^2 + (R_jm_i^{\star})^2}$ and $\chi_i$
the
quark spinor.  The effective quark mass, $m_i^{\star}$, is defined
by
\begin{equation}
m_i^{\star} = m_i - (V_{\sigma} \pm \frac{1}{2}V_{\delta}),
\mbox{ for a } {u \choose d} \mbox{quark}. \label{eq:qem}
\end{equation}
The linear boundary condition at the bag
surface provides the equation for the
eigenvalue $x_i$.
\par
Using the SU(6) spin-flavor nucleon wave function,
the nucleon energy is given by $E_{bag}^j + 3V_{\omega} \pm \frac{1}
{2}V_{\rho}$ for ${p \choose n }$, where the bag energy is
\begin{equation}
E_{bag}^j = {\frac{\sum_i n_{i/j} \Omega_{i/j} - z_0}{R_j}} +
{\frac{4}{3}}\pi
BR_j^3,
\label{eq:bageb}
\end{equation}
with $B$ the bag constant and $z_0$ a phenomenological parameter
(initially introduced to account for zero-point motion).
Here $n_{i/j}$ is the number of quarks of type $i$ in nucleon $j$.
To correct for spurious c.m. motion
in the bag~\cite{boost} the mass of the nucleon at rest is taken to be
\begin{equation}
M_j = \sqrt{(E_{bag}^j)^2 - \sum_i n_{i/j} (x_i/R_j)^2}.
\label{eq:cmc}
\end{equation}
The effective nucleon mass, $M_j^{\star}$, in nuclear matter is given
by minimizing eq.(\ref{eq:cmc}) with respect to $R_j$.
\par
To see the sensitivity of our results to the bag radius of the free
nucleon, we choose $m_u=5$ MeV and vary the parameters, $B$, $z_0$ and
$m_d$, to fit the physical proton and neutron masses for several
values of the average, free bag radius $R_0 (= 0.6, 0.8, 1.0 fm)$.
Since the
electromagnetic (EM) self-energies for {\it p\/} and {\it n\/}
contribute to the masses we adjust the parameters to fit the bare
proton mass, $M_p = 938.272-0.63$ MeV, and the bare neutron mass,
$M_n=939.566+0.13$ MeV, where $+0.63$ MeV and $-0.13$ MeV are the
EM self-energies for {\it p\/} and {\it n\/},
respectively~\cite{elms}.  We then find that $B^{1/4}$ = 187.4,
157.0, 135.9 MeV, $z_0$ = 2.043, 1.647, 1.178, and $m_d$ = 9.752,
10.01, 10.07 MeV for $R_0$ = 0.6, 0.8, 1.0 $fm$, respectively.  The
bag radius of the free proton is slightly smaller than that of the
free neutron, but the difference is negligible.
\par
For infinite nuclear matter we take the Fermi momenta for protons and
neutrons to be $k_{F_p}$ and $k_{F_n}$. These are
defined by $\rho_p = k_{F_p}^3 / (3\pi^3)$ and $\rho_n = k_{F_n}^3 /
(3\pi^3)$, where $\rho_p$ and $\rho_n$ are the densities of
{\it p\/} and {\it n}, respectively, and the total baryon density,
$\rho_B$, is given by $\rho_p + \rho_n$.  The $\omega$ field is
determined by baryon number conservation,
and the $\rho$ mean-field by the difference in proton and neutron
densities ($\rho_3$ below). On the other hand, the scalar mean-fields,
$\sigma$ and $\delta$, are given by a self-consistency
condition (SCC)~\cite{sath,serot}.  Since the $\rho$ field value is
given by $\bar{b} = g_{\rho} \rho_3 / (2m_{\rho}^2)$, where
$g_{\rho}=g_{\rho}^q$ and $\rho_3 = \rho_p - \rho_n$, the total
energy per nucleon, $E_{tot}$, can be written
\begin{equation}
E_{tot} = \frac{2}{\rho_B (2\pi)^3}\sum_{j=p,n}\int^{k_{F_j}}
d\vec{k} \sqrt{M_j^{\star 2} + \vec{k}^2} + \frac{m_{\sigma}^2}
{2\rho_B}{\bar{\sigma}}^2 + \frac{m_{\delta}^2}{2\rho_B}
{\bar{\delta}}^2 + \frac{g_{\omega}^2}
{2m_{\omega}^2}\rho_B + \frac{g_{\rho}^2}{8m_{\rho}^2\rho_B}
 \rho_3^2, \label{eq:toteb}
\end{equation}
where $g_{\omega} = 3g_{\omega}^q$.  Then, the SCC for the $\phi$
(= $\sigma$ or $\delta$) field is
\begin{equation}
\bar{\phi} = - \frac{2}{(2\pi)^3 m_{\phi}^2} \left[ \sum_{j=p,n}
\int^{k_{F_j}} d\vec{k} \frac{M_j^{\star}}
{\sqrt{M_j^{\star 2} + \vec{k}^2}} \times \left(\frac{\partial
M_j^{\star}}{\partial \bar{\phi}}\right)_{R_j} \right].
\label{eq:sccphi}
\end{equation}
Using eqs.(\ref{eq:bageb}) and (\ref{eq:cmc}), we find
\begin{equation}
\left(\frac{\partial M_p^{\star}}{\partial \bar{\sigma}}
\right)_{R_p} =
- g_{\sigma} \times ( 2C_{u/p}+C_{d/p})/3 \equiv - g_{\sigma} \times
C_p^{\sigma}, \label{eq:derps}
\end{equation}
\begin{equation}
\left(\frac{\partial M_n^{\star}}{\partial \bar{\sigma}}\right)_{R_n}
=
- g_{\sigma} \times (C_{u/n}+2C_{d/n})/3 \equiv - g_{\sigma} \times
C_n^{\sigma}, \label{eq:derns}
\end{equation}
\begin{equation}
\left(\frac{\partial M_p^{\star}}{\partial \bar{\delta}}\right)_{R_p}
=
\frac{g_{\delta}}{2} \times (-2C_{u/p}+C_{d/p}) \equiv
\frac{g_{\delta}}{2} \times C_p^{\delta}, \label{eq:derpd}
\end{equation}
\begin{equation}
\left(\frac{\partial M_n^{\star}}{\partial \bar{\delta}}\right)_{R_n}
=
\frac{g_{\delta}}{2} \times (-C_{u/n}+2C_{d/n}) \equiv
\frac{g_{\delta}}{2} \times C_n^{\delta}, \label{eq:dernd}
\end{equation}
where $g_{\sigma}=3g_{\sigma}^q$, $g_{\delta}=g_{\delta}^q$ and
$C_{i/j}$ is the scalar density of the {\it i}-th quark in {\it j}:
\begin{equation}
C_{i/j} = \left( \frac{E_{bag}^j}{M_j^{\star}} \right)
[(1-\frac{\Omega_{i/j}}{E_{bag}^jR_j}) S_{i/j} + \frac{m_i^{\star}}
{E_{bag}^j}], \label{eq:sd}
\end{equation}
with
\begin{equation}
S_{i/j} = \int d{\vec r} {\bar{\psi}}_{i/j}\psi_{i/j} =
\frac{\Omega_{i/j}/2+R_jm_i^{\star}(\Omega_{i/j}-1)}{\Omega_{i/j}
(\Omega_{i/j}-1)+R_jm_i^{\star}/2}. \label{eq:qsd}
\end{equation}
Note that while $C_p^{\delta}$ has a negative value the others are
positive, and that $C_{d/j}$ is slightly greater than $C_{u/j}$.
The reason for this is purely quantum mechanical: the {\it u\/} quark
is
{\it lighter, i.e., more relativistic,} than the {\it d\/} quark,
and hence the small Dirac component in the {\it u\/} quark wave
function is greater than that in the {\it d\/} quark.
In fig.1, we show the scalar density factors,
$C_j^{\phi}$, as functions of $\rho_B$.  The dependence of
the scalar densities on the bag radius is not strong.  Using eqs.
(\ref{eq:derps}) $\sim$ (\ref{eq:dernd}), the SCC for the $\phi$
field becomes
\begin{equation}
g_{\phi}\bar{\phi} = \frac{{2 \choose -1}}{(2\pi)^3} \left(
\frac{g_{\phi}}{m_{\phi}} \right) ^2 \left[ \sum_{j=p,n} C_j^{\phi}
\int^{k_{F_j}} d\vec{k} \frac{M_j^{\star}}{\sqrt{M_j^{\star 2} +
{\vec{k}}^2}} \right] , \mbox{ for } {\sigma \choose \delta}
\mbox{ meson.} \label{eq:sccsd}
\end{equation}
\par
We determine the coupling constants, $g_{\sigma}^2$ and
$g_{\omega}^2$, so as to fit the binding energy ($-16$ MeV) and
the saturation density ($\rho_0 = 0.17 fm^{-3}$) for equilibrium
nuclear matter. Furthermore, we choose $g_{\delta}^2/4\pi$ =
2.82~\cite{bonn} and fit the $\rho$ meson coupling constant so as
to reproduce the bulk symmetry energy of nuclear matter, $33.2$ MeV.
We then find that $g_{\sigma}^2/4\pi$ = 19.4, 20.8, 21.1,
$g_{\omega}^2/4\pi$ = 1.69, 1.26, 1.08, and $g_{\rho}^2/4\pi$ = 5.51,
5.74, 5.78 for $R_0$ = 0.6, 0.8, 1.0 {\it fm}, respectively.  (Here
we take $m_{\sigma}$ = 550 MeV, $m_{\omega}$ = 783 MeV, $m_{\rho}$ =
770 MeV and $m_{\delta}$ = 983 MeV.)  The present model gives a good
value for the nuclear compressibility -- around 220 MeV.
\par
Since we have seen the $\sigma$ meson field in Ref.\cite{sath},
here we show only the mean field values of the $\delta$ meson in
fig.2.  We define the proton fraction, $f_p$, as
$\rho_p/\rho_B$.  Clearly the $\delta$ meson field has a
symmetry property, ${\bar \delta}(f_p) = - {\bar \delta}(1-f_p)$
in the case where $m_u$ = $m_d$, so that in this case
$\bar{\delta}=0$ for $f_p$ = 0.5.  However, once the quark mass
difference is introduced this symmetry is lost; even for $f_p$ = 0.5
the $\delta$ field does not vanish but has a negative value.  It is
easy to see why it is negative: from eq.(\ref{eq:sccsd}), the
$\sigma$ and $\delta$ fields at low $\rho_B$ can be expressed as
\begin{equation}
g_{\sigma}\bar{\sigma} \simeq \left( \frac{g_{\sigma}}{m_{\sigma}}
\right) ^2 ( C_p^{\sigma} \rho_p + C_n^{\sigma} \rho_n),
\label{eq:slow}
\end{equation}
\begin{equation}
g_{\delta}\bar{\delta} \simeq - \frac{1}{2} \left( \frac{g_{\delta}}
{m_{\delta}} \right) ^2 ( C_p^{\delta} \rho_p + C_n^{\delta}
\rho_n ). \label{eq:dlow}
\end{equation}
Therefore, when $f_p$ = 0.5, $g_{\delta}\bar{\delta} = - \frac{1}{4}
\left( \frac{g_{\delta}}{m_{\delta}} \right) ^2 ( C_p^{\delta} +
C_n^{\delta} ) \rho_B $ which is less than zero because, as
explained below eq.(\ref{eq:qsd}), $ C_p^{\delta} + C_n^{\delta} > 0$.
\par
At this stage we are ready to consider the NS anomaly.
As the first-order approximation,
the binding energy difference, eq.(\ref{eq:def}), would be treated in
terms of the difference between the effective masses of {\it p\/} and
{\it n\/} in matter: $\Delta E_c = \Delta_{np}^0 - ( M_n^{\star} -
M_p^{\star} )$, where $\Delta_{np}^0$ = 1.293 MeV is the {\it n-p\/}
mass difference in free space.  Since the effective masses,
$M_p^{\star}$ and $M_n^{\star}$, are functions of $\rho_B$
in our model, we may use the local density approximation (LDA) to
calculate $\Delta E_c$ for finite nuclei.  For medium mass nuclei,
$\Delta E_c$ has been measured using mirror nuclei~\cite{shlomo}.
On the other hand, for large nuclei ({\it i.e.}, neutron rich nuclei),
the experiments have been performed by measuring analog resonances in
$(p,p^{\prime})$ and $(p,n)$ reactions~\cite{seitz}.  In LDA the
effective {\it n-p\/} mass difference in the nucleus A,
$\Delta_{np/A}^{\star}$, may be calculated as
\begin{equation}
\Delta_{np/A}^{\star} = \int d\vec{r} \Delta_{np}^{\star}(\bar{\rho},
f_p) P(r), \label{eq:ldaa}
\end{equation}
where $\bar{\rho}$ describes the core density of the mirror nuclei or
the nuclear density for large mass nuclei, which is
taken to have a Woods-Saxon form: ${\bar{\rho}}_0 / (1 +
\exp[\frac{r-\bar{R}}{a}])$ with ${\bar{\rho}}_0$ the normalization
factor and $a \simeq 0.54 fm$.  Here we take $\bar{R}$ so as to
reproduce the rms radius of the core nucleus for medium mass
nuclei~\cite{wilth} or $\bar{R} \simeq 1.12(fm)A^{1/3} -
0.86(fm)A^{-1/3}$ for neutron rich nuclei.  Furthermore, we choose
$\Delta_{np}^{\star} = M_n^{\star}(\bar{\rho}, f_p) - M_p^{\star}
(\bar{\rho}, f_p)$ with $f_p=0.5 (Z/A)$ for mirror (large) nuclei. The
probability distribution, $P(r)$, of the last nucleon in mirror
nuclei is given by the harmonic oscillator model with the radius
parameter chosen to fit the rms radius of the valence
nucleon~\cite{wilth}. (For large nuclei we use the nuclear density
distribution itself for $P(r)$.)
We show the bag radius dependence of $\Delta_{np}^{\star}$ in
fig.3.
It gradually decreases as $\rho_B$ grows, becoming negative for
high densities, and it does not much depend on $R_0$.  We can
easily see why it decreases: since the
effective nucleon mass can be expanded at low densities as
\begin{equation}
M_j^{\star} \simeq M_j - g_{\sigma} C_j^{\sigma} \bar{\sigma} +
\frac{g_{\delta}}{2} C_j^{\delta} \bar{\delta}, \label{eq:efmas}
\end{equation}
we find that $\Delta_{np}^{\star} = \Delta_{np}^0 - g_{\sigma}
(C_n^{\sigma} - C_p^{\sigma}) \bar{\sigma} + \frac{1}{2} g_{\delta}
(C_n^{\delta} - C_p^{\delta}) \bar{\delta}$ when $f_p$ = 0.5.
Because $C_n^{\sigma} > C_p^{\sigma}$ and $\bar{\delta}$ is
numerically much smaller than $\bar{\sigma}$, the {\it n-p\/} mass
difference decreases as $\rho_B$ goes up.
The numerical results for $\Delta E_c ( = \Delta_{np}^0 -
\Delta_{np/A}^{\star} )$ are displayed in table~1.  The
present model provides a good overall description of the
variation of $\Delta E_c$ with mass number.  We can see the
noticeable shell effects arising from the differences in the average
density felt by the valence nucleon.  The calculated values are of
the right order of magnitude to resolve the anomaly.
Note, however, that a complete treatment of the anomaly, including the
spin-orbit effects, goes beyond the local density approach used here.
\par
Next we wish to use the model to study the quark condensates in
matter.
Within the QCD sum-rule approach these
play a very important role in a wide range of nuclear
phenomena.  The difference between the {\it i}-th quark condensates
in matter, $Q_i(\rho_B)$, and in vacuum, $Q_i(0)$, can be related to
the meson fields through the Hellmann-Feynman
theorem~\cite{sath,cohen}:
\begin{eqnarray}
{\tilde Q}_i(\rho_B) \equiv Q_i(\rho_B) - Q_i(0) &=& \sum_{j=p,n}
\frac{\partial {\cal E}}{\partial M_j^{\star}} \frac{\partial
M_j^{\star}}{\partial m_i} + (\mbox{meson contributions}),
\nonumber \\
 &=& n_{i/p} C_{i/p} \frac{\partial {\cal E}}{\partial
M_p^{\star}} + n_{i/n} C_{i/n}
 \frac{\partial {\cal E}}{\partial
M_n^{\star}} + {\cal O}(\rho_B^2), \label{eq:cond}
\end{eqnarray}
where ${\cal E} = \rho_B E_{tot}$ and the meson fields contribute
${\cal O}(\rho_B^2)$.  We take up the terms of ${\cal O}(\rho_B)$ to
see the quark condensate at low density.  Defining the average of the
quark condensates, ${\cal Q}$, by $\frac{1}{2}({\tilde Q}_u +
{\tilde Q}_d)$ and their difference, $\delta {\cal Q}$, by
${\tilde Q}_d - {\tilde Q}_u$, they can be expressed as
\begin{equation}
{\cal Q}(\rho_B) \simeq \frac{3}{2} \left( \frac{m_{\sigma}}
{g_{\sigma}} \right) ^2 (g_{\sigma}\bar{\sigma}) \simeq \frac{3}{2}
(C_p^{\sigma} \rho_p + C_n^{\sigma} \rho_n), \label{eq:avqcon}
\end{equation}
and
\begin{equation}
\delta {\cal Q}(\rho_B) \simeq - 2 \left( \frac{m_{\delta}}
{g_{\delta}} \right) ^2 (g_{\delta} \bar{\delta}) \simeq C_p^{\delta}
\rho_p + C_n^{\delta} \rho_n, \label{eq:difqcon}
\end{equation}
where we used eqs.(\ref{eq:sccsd}) $\sim$ (\ref{eq:dlow}).  It is
worth noting that at low density ${\cal Q}$ and $\delta {\cal Q}$
can be described by the $\sigma$ and $\delta$ fields
(respectively) {\it alone}.  Since we have already discussed
${\cal Q}$ in
Ref.\cite{sath}, we shall concentrate on $\delta {\cal Q}$.
In particular it can be related
to $\gamma$, which is a measure of isospin symmetry
breaking in the quark condensate in matter, defined by $\gamma =
Q_d(\rho_B)/Q_u(\rho_B) -1$.  Using eqs.(\ref{eq:dlow}) and
(\ref{eq:difqcon}), we find that at low density $\gamma$ is
\begin{equation}
\gamma \simeq \gamma_0 + \frac{\delta {\cal Q}}{Q_u(0)} \simeq
\gamma_0 + \frac{2}{\vert Q_u(0) \vert} \left( \frac{m_{\delta}}
{g_{\delta}} \right) ^2 (g_{\delta} \bar{\delta}) \simeq \gamma_0 -
\frac{1}{\vert Q_u(0) \vert} ( C_p^{\delta} \rho_p + C_n^{\delta}
\rho_n), \label{eq:gamma}
\end{equation}
where $\vert Q_u(0) \vert \simeq (225 \mbox{MeV})^3$~\cite{qcdsum}
and $\gamma_0$ is the value of $\gamma$ in vacuum: $\gamma_0 = -
(6 \sim 9) \times 10^{-3}$~\cite{hb,qcdsum}.  Therefore, one finds
that $\gamma \simeq \gamma_0 - 1.0 (1.3) \times 10^{-3} (\rho_B /
\rho_0)$ for $R_0 = 0.6 (1.0) fm$ when $f_p$ = 0.5.  Since $\gamma$
is directly related to $\bar{\delta}$, $\gamma$ depends not only on
$\rho_B$ but also on $f_p$, which is quite different from the result
in the NJL model~\cite{hb}.  Our result, however, seems natural
because the origin of isospin symmetry breaking is very clear in our
model: the $\delta$ meson couples {\it differently\/} to the {\it u\/}
and {\it d\/} quarks and the difference of their effective masses in
matter
is completely controlled by $\bar{\delta}$
(see eq.(\ref{eq:qem})).  Note that $\gamma$ can be rewritten in
terms of the quark effective mass: $\gamma \simeq \gamma_0 + \frac{2}
{\vert Q_u(0) \vert} ( \frac{m_{\delta}}{g_{\delta}} ) ^2
( \delta m_d^{\star} - \delta m_u^{\star})$, where $\delta
m_i^{\star} = m_i^{\star} -m_i$.
\par
Finally, we shall consider a  potential experimental signature
of isospin symmetry
breaking.  Since the {\it u-d\/} quark mass difference leads a
difference between small Dirac components in {\it u\/} and {\it d\/}
quark wave functions, the axial-vector neutral-current coupling
constants of {\it p\/} and {\it n\/}, $g_A^p$ and $g_A^n$, seem
interesting~\cite{tbg}.  Using eq.(\ref{eq:dirac}), the ratio of
$g_A^n$ to $g_A^p$ is given by $g_A^n/g_A^p = (4G_{d/n}+G_{u/n})/
(4G_{u/p}+G_{d/p})$, where $G_{i/j}= {\cal N}_{i/j}^2 \int dr r^2
[ j_0^2(x_i r/R_j) - \frac{1}{3} \beta_{i/j}^2 j_1^2(x_i r/R_j)] =
(2\Omega_{i/j}^2+4\Omega_{i/j}R_jm_j^{\star}-3R_jm_i^{\star})/3
(2\Omega_{i/j}(\Omega_{i/j}-1)+R_jm_i^{\star})$.
In symmetric nuclear matter, we find that $g_A^n/g_A^p = 1 + \frac{3}
{5}\eta$ with $\eta \simeq 2 (4) \times 10^{-3} + 0.2 (6) \times
10^{-4} (\rho_B/\rho_0)$ for $R_0 = 0.6 (1.0) fm$.  The dependence of
$\eta$ on $\rho_B$ is quite weak.  At the present experimental level
it is not possible to see it, but one may hope in the future.
\par
In summary, we have applied the QMC model involving $\sigma$,
$\omega$, $\rho$ and $\delta$ mesons (all mesons whose masses lie
below
1 GeV) to investigate the NS anomaly for finite nuclei. We have also
used it to investigate isospin
symmetry breaking in the quark condensate in nuclear matter.
The result is a very reasonable
description of both the anomaly and $\gamma$ in nuclear matter.
Their physical origin is very clear and the differences
between the quark scalar densities in the proton and neutron, which
are
generated by the {\it u-d\/} quark mass difference, are essential
to understand them.
\par
This work was supported by the Australian Research Council.
%
%

%
\newpage
\begin{flushleft}
\Large{Figure captions}
\end{flushleft}
\begin{description}
\item[Fig.1] Scalar density factors, $C_j^{\phi}$, in symmetric nuclear
matter ($R_0 = 0.8 fm$).  The solid, dotted, dashed and dot-dashed
curves are for $C_p^{\sigma}$, $C_n^{\sigma}$,
$\vert C_p^{\delta}\vert$ and $C_n^{\delta}$, respectively.  The
values of the coupling constants are given in the text.
\item[Fig.2] Mean field values of the $\delta$ meson for $R_0 = 0.8 fm$.
The dotted curves are for the case where $m_u$ = $m_d$ = 10 MeV.  The
solid curves display the full calculation involving the quark mass
difference.  The top (middle) [bottom] two curves are for $f_p$ = 0.7
(0.5) [0.3].
\item[Fig.3] The {\it n-p\/} mass difference in matter
($f_p = 0.5$).  The dotted,
solid and dashed curves are for the calculations with $R_0$ =
0.6, 0.8 and 1.0 $fm$, respectively.
\end{description}
\newpage
\begin{table}
\begin{center}
\caption{$\Delta E_c$ for several finite nuclei.  Energies are
quoted in MeV.}
\label{nsaf}
\begin{tabular}[t]{ccccc}
\hline
 $R_0 (fm)$ & 0.6 & 0.8 & 1.0 & observed discrepancy \\
\hline
 $^{15}$O--$^{17}$N & 0.29 & 0.32 & 0.34 & 0.16 $\pm$ 0.04 \\
 $^{17}$F--$^{17}$O & 0.22 & 0.25 & 0.27 & 0.31 $\pm$ 0.06 \\
 $^{39}$Ca--$^{39}$K & 0.36 & 0.41 & 0.44 & 0.22 $\pm$ 0.08 \\
 $^{41}$Sc--$^{41}$Ca & 0.34 & 0.38 & 0.41 & 0.59 $\pm$ 0.10  \\
 $^{120}$Sn & 0.72 & 0.83 & 0.87 &  \\
 $^{208}$Pb & 0.78 & 0.91 & 0.95 & $\sim$ 0.9 \\
\hline
\end{tabular}
\end{center}
\end{table}
\end{document}